\documentclass[12pt]{article}
\usepackage{a4wide}
\title{Bitcoin Returns and Public Attention to COVID-19: Do Timing and Individualism Matter?\thanks{Declaration of Interest: None.}}
\author{Huaxin Wang-Lu\thanks{Corresponding author. Department of Quantitative Methods, IQS School of Management, Universitat Ramon Llull, Via Augusta 390, 08017 Barcelona, Spain. Tel.: +34 637 170 126. E-mail: \color{blue}{huaxin.wanglu@iqs.url.edu.}}}
\date{}
\usepackage{microtype}
\usepackage[margin=2.5cm]{geometry}
\usepackage{mathpazo}
\usepackage{antpolt}
\usepackage[T1]{fontenc}
\usepackage{libertine}
\usepackage[section]{placeins} 
\usepackage{anyfontsize}
\usepackage[euler]{textgreek}
\usepackage{amsmath}
\usepackage{lipsum} 
\usepackage{wasysym} 
\usepackage[nodisplayskipstretch]{setspace} 
\usepackage{graphicx} 
\usepackage{stfloats} 
\usepackage{subcaption} 
\usepackage{longtable} 
\usepackage{supertabular}
\usepackage{multirow} 
\usepackage{ragged2e,array,booktabs} 
\usepackage{threeparttablex}
\usepackage[table]{xcolor} 
\usepackage{mwe}
\usepackage{booktabs} 
\usepackage[normalem]{ulem} 
\usepackage{caption} 
\usepackage{threeparttable}  
\usepackage[style=authoryear-icomp,doi=false,isbn=false,url=false,eprint=false,backend=biber,autocite=inline, labeldateparts=true, uniquename=false, uniquelist=false, ibidtracker=false, pagetracker=true, sortcites=true]{biblatex}

\DeclareNameWrapperFormat{labelname:poss}{#1's}

\newrobustcmd*{\posscitealias}{%
  \AtNextCite{%
    \DeclareNameWrapperAlias{labelname}{labelname:poss}}}

\newrobustcmd*{\posscite}{%
  \posscitealias
  \textcite}

\newrobustcmd*{\Posscite}{\bibsentence\posscite}

\newrobustcmd*{\posscites}{%
  \posscitealias
  \textcites}

\usepackage[hypertexnames=false,citecolor=blue,colorlinks=true,linkcolor=blue,urlcolor=green,bookmarksopen=true]{hyperref} 

\usepackage{bookmark}

\graphicspath{{/Users/huaxinwanglu/Library/Mobile Documents/com~apple~CloudDocs/Bitcoin Paper/Latex/}}

\DeclareFieldFormat{citehyperref}{%
  \DeclareFieldAlias{bibhyperref}{noformat}
  \bibhyperref{#1}}

\DeclareFieldFormat{textcitehyperref}{%
  \DeclareFieldAlias{bibhyperref}{noformat}
  \bibhyperref{%
    #1%
    \ifbool{cbx:parens}
      {\bibcloseparen\global\boolfalse{cbx:parens}}
      {}}}

\savebibmacro{cite}
\savebibmacro{textcite}

\renewbibmacro*{cite}{%
  \printtext[citehyperref]{%
    \restorebibmacro{cite}%
    \usebibmacro{cite}}}

\renewbibmacro*{textcite}{%
  \ifboolexpr{
    ( not test {\iffieldundef{prenote}} and
      test {\ifnumequal{\value{citecount}}{1}} )
    or
    ( not test {\iffieldundef{postnote}} and
      test {\ifnumequal{\value{citecount}}{\value{citetotal}}} )
  }
    {\DeclareFieldAlias{textcitehyperref}{noformat}}
    {}%
  \printtext[textcitehyperref]{%
    \restorebibmacro{textcite}%
    \usebibmacro{textcite}}}

\begin{document}

\maketitle

\begin{abstract}
\singlespacing
\noindent 
The evolution of the pandemic and people's concern over it have an impact on the Bitcoin market, while the extent of individualism could differentiate investor behaviors in the financial market during the pandemic. This paper examines whether public attention to COVID-19  in individualistic countries versus collectivistic countries Granger causes Bitcoin returns between February 11, 2020 and May 09, 2022. To this end, eight large economies with a individualistic or collectivistic tradition are chosen for analyses. By using rolling and recursive-evolving algorithms, it accounts for the timing of COVID-19 issues that vary by country and circumvents the potential estimation bias that a traditional Granger causality test may suffer due largely to Google's sampling variation for different time frames. In general, collectivistic countries are found to have stronger causal impacts on Bitcoin returns than individualistic countries.

\medskip
\hspace{-1.4em}\textbf{JEL Classification:} G41; G12; G15

\hspace{-1.4em}\textbf{Keywords:} Bitcoin; COVID-19; individualism; Google Trends; time-varying causality

\end{abstract}

\newpage

\section{Introduction}
\hspace{1.4em}Since the first known outbreak of COVID-19, Bitcoin price, at its peak, skyrocketed by more than 700\%. There is evidence that the price explosion coincided with mortality caused by COVID-19 \autocite{Goodell2021}, but not only deaths, confirmed cases and news related to COVID-19 were also correlated with the bitcoin market \autocite{Sarkodie2022, Zhang2022}. The evolution, particularly, a worsening prospect, of the pandemic and the exposure to its news could effectively stir up public panic. These fears towards COVID-19 were often accompanied by a drop in yield at hand, but a reversal later, and boosted transaction volume and volatility \autocite{Chen2020a}.\par


\hspace{1.4em}At the same time, the epidemic situation varies widely across countries due, in part, to the role of cross-cultural differences in preventive behaviors and health outcomes. More individualistic regions saw less compliance with stay-at-home orders, were more reluctant to a variety of socially optimal actions, such as social distancing, mask use, or receiving vaccine, and had more cases and/or higher fatalities (e.g., \cites{Maaravi2021, Bazzi2021}). 

\hspace{1.4em}Overall, the pandemic and people's concern over it form part of the dynamics of the Bitcoin market, while individualism and collectivism, as a cultural element, differentiate how individuals regard and react to the viral spread and corresponding measures. Thus, do individualistic and collectivistic countries differ in their influence on the Bitcoin market under the the shock of COVID-19? Despite indirectly, possible answers could be drawn from the previous studies that increases in individualism was correlated with decreases in Bitcoin price co-movements \autocite{Caporale2020}; individualistic culture weakened the willingness to buy Bitcoin during the lockdown \autocite{Chen2022}. However, people's attention and the ensuing panic depends on when COVID-19 and its variants occur and spread, particularly in their countries. This means that time is crucial to the analysis.\footnote{In addition, both the fact that testing for Granger causality is sensitive to the sample period and the underlying assumption of unchanged causality over time point towards the necessity of examining the time-varying causality in this context.}\par


\hspace{1.4em}Hence, this paper analyzes the time-varying Granger causality between public attention to COVID-19 in individualistic countries versus collectivistic countries and Bitcoin returns on a daily basis. To this end, two newly-developed time-varying Granger causality tests are adopted \autocites{Shi2018}, and Google Search Volume Index (GSVI) data are exploited.\footnote{They are normalized indices reflecting real-time interest in a topic relative to all topic searches on Google. It allows researchers to compare trends across countries and dates, independent of the disparity in the user population.} Because it is impossible to analyze every country on a case-by-case basis, I chose as a case study the United States, China, Japan, Germany, the United Kingdom, India, France and South Korea.\footnote{They are both the largest economies in the world and traditionally individualistic (collectivistic) countries.} In addition, GSVI data are inconsistent across time due largely to the variation in random samples drawn by Google for different time frames \autocite{Eichenauer2022}, which could invalidate a traditional Granger causality test. With time-varying causality tests, I circumvent this issue by controlling the initialization and bootstrapping procedure to be performed over a period of three months.\footnote{I used a three-month time frame to download all the GSVI data.}\par


\section{Data and Methodology}

\subsection{Data Source}

\hspace{1.4em}Daily data on Bitcoin prices from February 11, 2020 to May 9, 2022 are retrieved from \href{www.bitcoincharts.com}{Bitcoincharts}.\footnote{The World Health Organization released the official name ``COVID-19'' on February 11, 2020.} Bitcoin return is defined as $r_{t}=\ln(\frac{P_{t}}{P_{t-1}})$. Daily attention to COVID-19 issues is measured by regional Google Search Volume Index, using the keyword ``COVID''.\footnote{Using ``COVID'' leads to more inclusive results than using ``COVID-19''. See \href{ https://support.google.com/trends/answer/4359582?hl=en}{search tips} here.} The data are briefly summarized in Table 1.\par

\begin{table}[htp]
\captionsetup{belowskip=0pt,aboveskip=1pt}
\noindent\makebox[\textwidth]{
\fontsize{11}{12}\selectfont{
\begin{threeparttable}
\caption{Descriptive Statistics\label{tab8}}
\begin{tabular}{l c c}    
\toprule
Country & Daily GSVI Mean & Individualism Score \\
\midrule
\textit{Individualism} \\
\midrule
US & 58.68 (2nd) & 91 (highest) \\
UK & 55.45 (4th) & 89 (2 lower the US) \\
France & 49.55 (6th) & 71 (20 lower the US) \\
Germany & 65.88 (1st) & 67 (24 lower the US) \\
\midrule
\textit{Collectivism} \\
\midrule
South Korea & 56.26  (3rd) & 18 (lowest) \\
China & 42.42 (8th) & 20 (2 higher than South Korea) \\
Japan & 50.97 (5th) & 46 (28 higher than South Korea) \\
India & 43.72 (7th) & 48 (30 higher than South Korea) \\
\bottomrule
\end{tabular}
\footnotesize{Notes: The scale of GSVI and individualism score is 0--100 and 6--91, respectively. \\ Source: Google Trends \& \textcite{Hofstede2010}.}
\end{threeparttable}}
}
\end{table}

\FloatBarrier

\subsection{Time-Varying Granger Causality Tests}

\hspace{1.4em}Following \posscite{Shi2018} procedure, an unrestricted VAR(p) model can be written as:
\begin{align}
y_{t}=\phi_{0}+\phi_{i}\sum_{i=1}^{p} y_{t-i}+\epsilon_{t}
\end{align}
where $y_{t}$ is a vector of variables of interest. $\phi_{0}$ is a vector of constants, and $\epsilon_{t}$ is a vector of independent white noise innovations.\par

\hspace{1.4em}The Wald test of the restrictions imposed by the null hypothesis has the general form:
\begin{equation}
W=[\textbf{R}\: \operatorname{vce}(\hat{\Pi})]'[\textbf{R}(\hat{\Omega}\otimes (X'X)^{-1})\textbf{R}']^{-1}[\textbf{R}\: \operatorname{vce}(\hat{\Pi})]
\end{equation}
where $\operatorname{vce}(\hat{\Pi})$ denotes the (row vectorized) $2(2p+1)\times 1$ coefficients of $\hat{\Pi}$, and $\textbf{R}$ is the $p\times 2(2p+1)$ selection matrix. Each row of $\textbf{R}$ picks one of the coefficients to set to zero under the non-causal null hypothesis.\par

\hspace{1.4em}\textcite{Shi2018} proposes three tests based on the supremum norm (sup) of a series of recursively evolving Wald test statistics to detect changes in causality.\footnote{The power of the recursive evolving procedure is found to be best, followed by the rolling procedure.} The origination (termination) date of a change in causality is identified as the first observation whose test statistic value exceeds (goes below) its corresponding critical value.\par


\hspace{1.4em}The Wald statistic obtained for each sub-sample regression, using observations over $[f_{1}, f_{2}]$ with a sample size fraction of $f_{w}=f_{2}-f_{1} \geq f_{0}$, is denoted by $W_{f_{2}}(f_{1})$, and the supremum Wald statistic is defined as:
\begin{equation}
\operatorname{SW}_{f}(f_{0})=\underset{(f_{1}, f_{2})\in \Lambda_{0}, f_{2}=f}{\operatorname{sup}}\left \{W_{f_{2}}(f_{1})\right \}
\end{equation}
where $\Lambda_{0}=\{(f_{1}, f_{2}): 0<f_{0}+f_{1}\leq f_{2} \leq 1, and \ 0\leq f_{1} \leq 1-f_{0}\}$ for some minimal sample size $f_{0}\in (0,1)$ in the regressions. This is the so-called recursive evolving procedure.\par

\hspace{1.4em}Let $f_{e}$ and $f_{f}$ denote the origination and termination points in the causal relationship that are estimated as the first chronological observation whose test statistic exceeds or falls below the critical value. The analysis relies on dating rules of the following two algorithms:
\begin{align}
\begin{split}
\text{Rolling}: \ \hat{f_{e}}=\underset{f\in [f_{0},1]}{inf} \{f: W_{f}(f-f_{0})>\operatorname{cv}\} \ \text{and} \ \hat{f_{f}}=\underset{f\in [\hat{f_{e}},1]}{inf}\{f: W_{f}(f-f_{0})<\operatorname{cv}\} \\
\text{Recursive Evolving}: \ \hat{f_{e}}=\underset{f\in [f_{0},1]}{inf} \{f: \operatorname{SW}_{f}(f_{0})>\operatorname{scv}\} \ \text{and} \ \hat{f_{f}}=\underset{f\in [\hat{f_{e}},1]}{inf} \{f: \operatorname{SW}_{f}(f_{0})<\operatorname{scv}\}
\end{split}
\end{align}
where $\operatorname{cv}$ and $\operatorname{scv}$ are the corresponding critical values of the $W_{f}$ and $\operatorname{SW}_{f}$ statistics. All procedures are implemented under the null hypothesis of no causality and under the assumption of either homoskedasticity or conditional heteroskedasticity of an unknown form.\par


\section{Empirical Results}

\hspace{1.4em}To estimate the VAR model and implement the aforementioned tests, the Bayesian information criteria (BIC) with a maximum potential lag length of 12 are used to select the lag order. The minimum window size $f_{0}$ is set to be three months. Critical values are obtained from a bootstrapping procedure with 499 replications. The empirical size is 5\% and controlled over a three-month period. Further, the Augmented Dickey-Fuller (ADF) and Phillips-Perron unit root tests suggest that Bitcoin returns and daily attention are all stationary (p-value $< 1\%$ for all).\par


\subsection{Individualistic Countries}



\hspace{1.4em}The results of the recursive evolving tests running from the public attention to COVID-19 in the US, Germany, the UK, and France to Bitcoin returns are displayed in Figures 1--4.\footnote{Test results using rolling window algorithms are reported in the Appendix.} As seen in Figure 1, two Wald tests detect that no statistic sequence exceeds its corresponding critical values, indicating that the null hypothesis of no Granger causality cannot be rejected for the U.S. Whereas, in Figure 2, a very short causal episode is identified for the UK. Further, the results for Germany are inconsistent -- no causality is observed in Figure 3(a), but two causal episodes are found in Figure 3(b).  More importantly, the attention paid to the pandemic by people living in France had an impact on Bitcoin returns during at least two periods, among which the latest one arose on January 22, 2022 and terminated on February 14, 2022. In general, no sufficient evidence can be drawn that people's concern over the epidemic situations in these individualistic countries has an impact on Bitcoin returns.\par 


\begin{figure}[thp]
  \begin{subfigure}[b]{0.5\textwidth}
  \centering
    \includegraphics[scale=0.5]{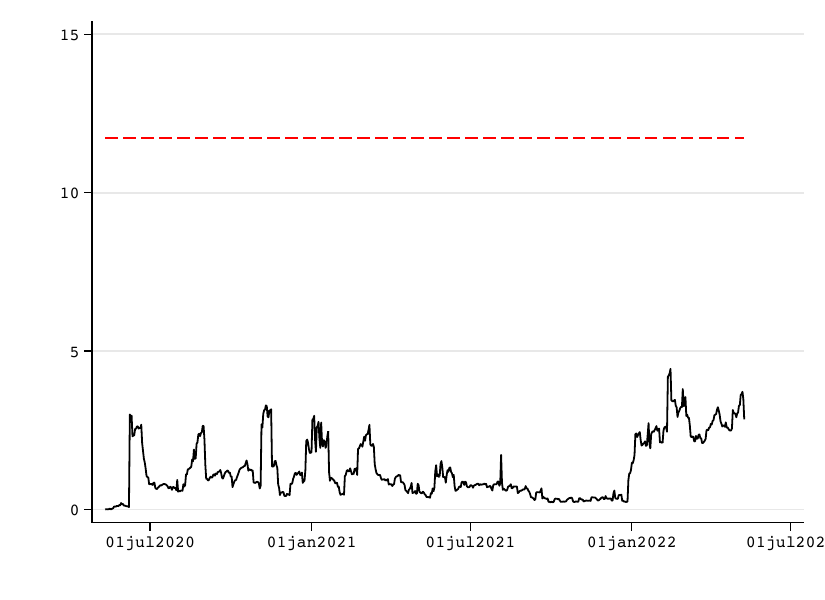}
    \caption{\fontsize{10}{12}\selectfont{Recursive Evolving-Homoskedasticity.}}
  \end{subfigure}
  \hfill
  \begin{subfigure}[b]{0.5\textwidth}
  \centering
    \includegraphics[scale=0.5]{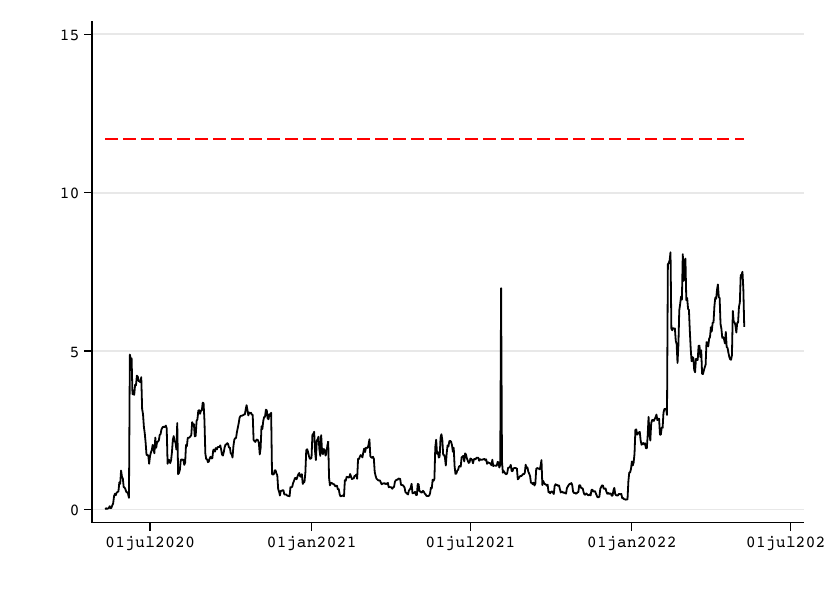}
    \caption{\fontsize{10}{12}\selectfont{Recursive Evolving-Heteroskedasticity.}}
  \end{subfigure}
  \caption{\fontsize{10}{12}\selectfont{Causality tests for attention to COVID-19 in the US --> Bitcoin returns. \\ Notes: The test statistic sequence (---) is in black; the 5\% bootstrapped critical value sequence (--) is in red. The selected lag order is 1.}}
\end{figure}

\begin{figure}[thp]
  \begin{subfigure}[b]{0.5\textwidth}
  \centering
    \includegraphics[scale=0.5]{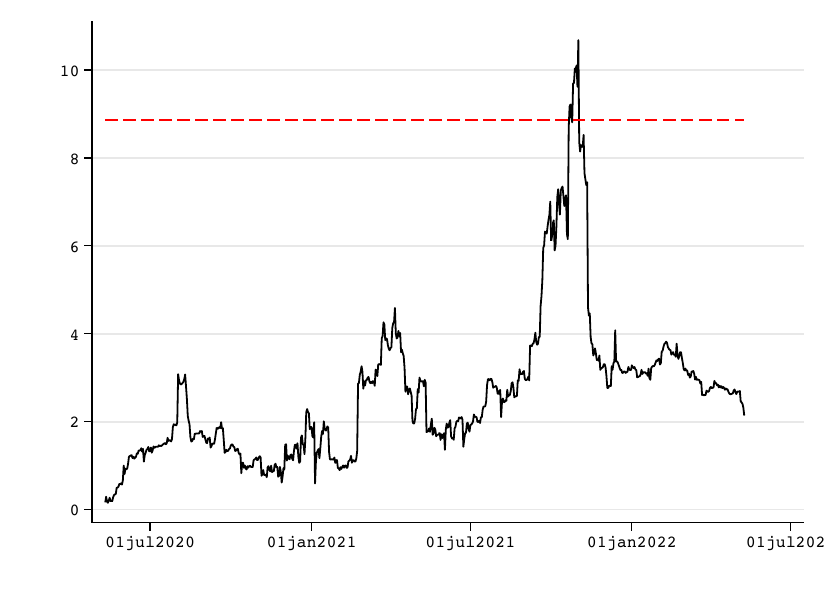}
    \caption{\fontsize{10}{12}\selectfont{Recursive Evolving-Homoskedasticity.}}
  \end{subfigure}
  \hfill
  \begin{subfigure}[b]{0.5\textwidth}
  \centering
    \includegraphics[scale=0.5]{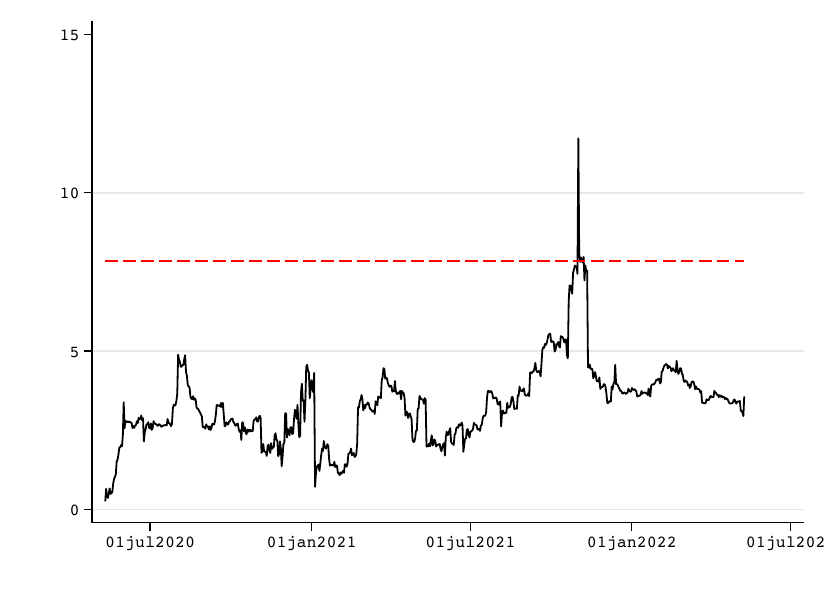}
    \caption{\fontsize{10}{12}\selectfont{Recursive Evolving-Heteroskedasticity.}}
  \end{subfigure}
  \caption{\fontsize{10}{12}\selectfont{Causality tests for attention to COVID-19 in the UK --> Bitcoin returns. \\ Notes: The test statistic sequence (---) is in black; the 5\% bootstrapped critical value sequence (--) is in red. The selected lag order is 1.}}
\end{figure}

\begin{figure}[thp]
  \begin{subfigure}[b]{0.5\textwidth}
  \centering
    \includegraphics[scale=0.5]{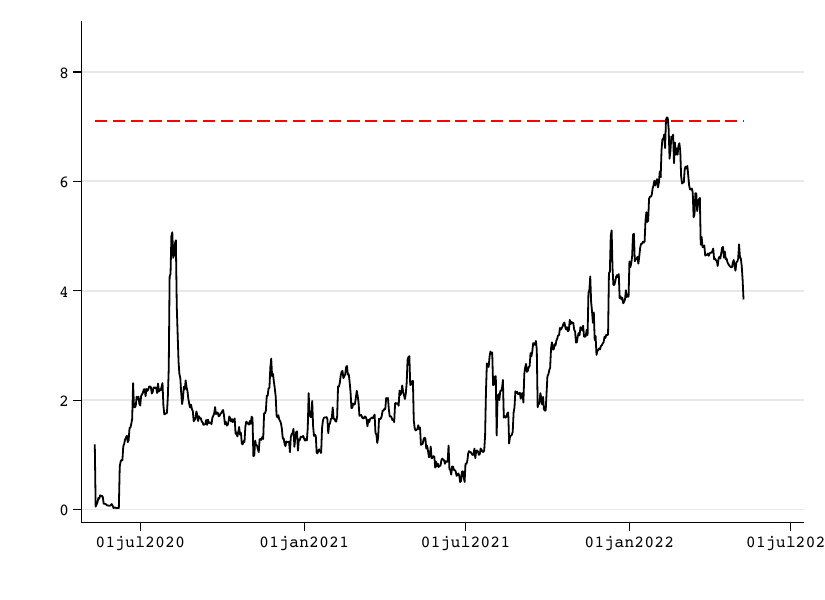}
    \caption{\fontsize{10}{12}\selectfont{Recursive Evolving-Homoskedasticity.}}
  \end{subfigure}
  \hfill
  \begin{subfigure}[b]{0.5\textwidth}
  \centering
    \includegraphics[scale=0.5]{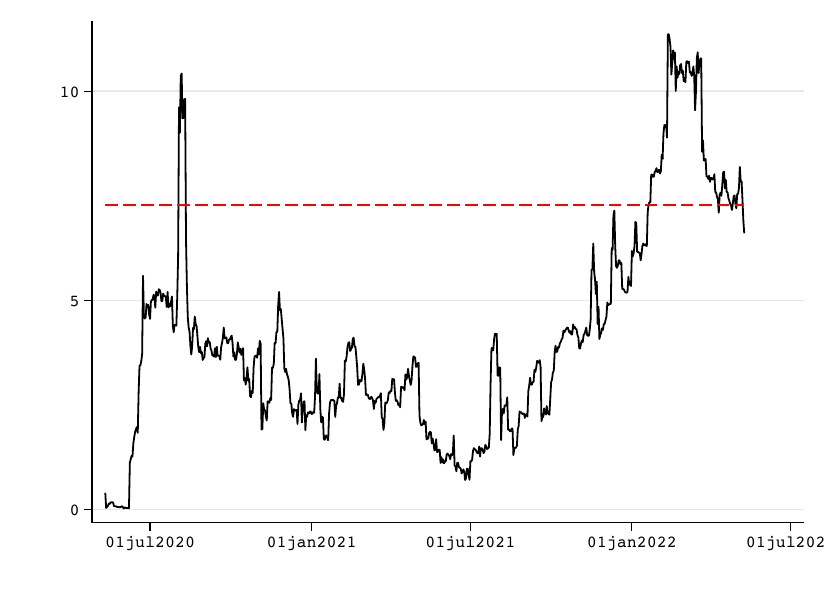}
    \caption{\fontsize{10}{12}\selectfont{Recursive Evolving-Heteroskedasticity.}}
  \end{subfigure}
  \caption{\fontsize{10}{12}\selectfont{Causality tests for attention to COVID-19 in Germany --> Bitcoin returns. \\ Notes: The test statistic sequence (---) is in black; the 5\% bootstrapped critical value sequence (--) is in red. The selected lag order is 1.}}
\end{figure}

\begin{figure}[thp]
  \begin{subfigure}[b]{0.5\textwidth}
  \centering
    \includegraphics[scale=0.5]{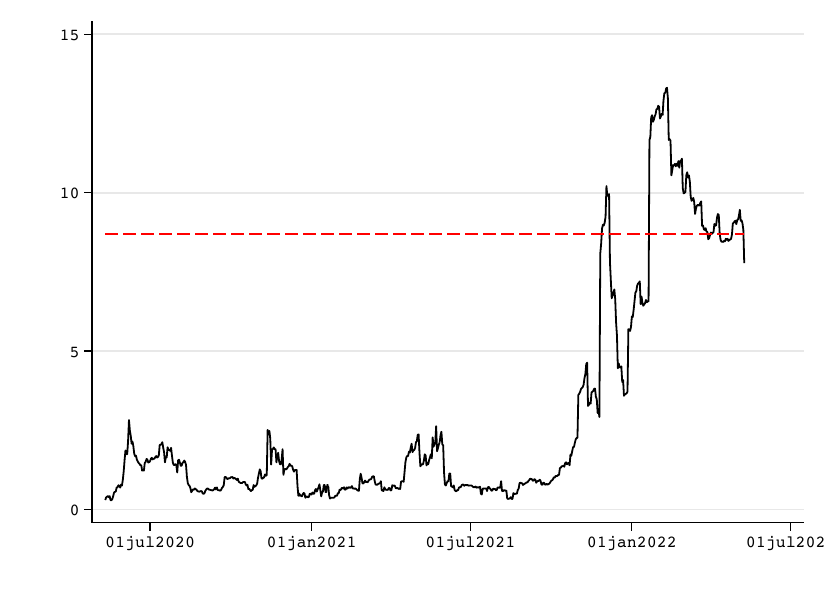}
    \caption{\fontsize{10}{12}\selectfont{Recursive Evolving-Homoskedasticity.}}
  \end{subfigure}
  \hfill
  \begin{subfigure}[b]{0.5\textwidth}
  \centering
    \includegraphics[scale=0.5]{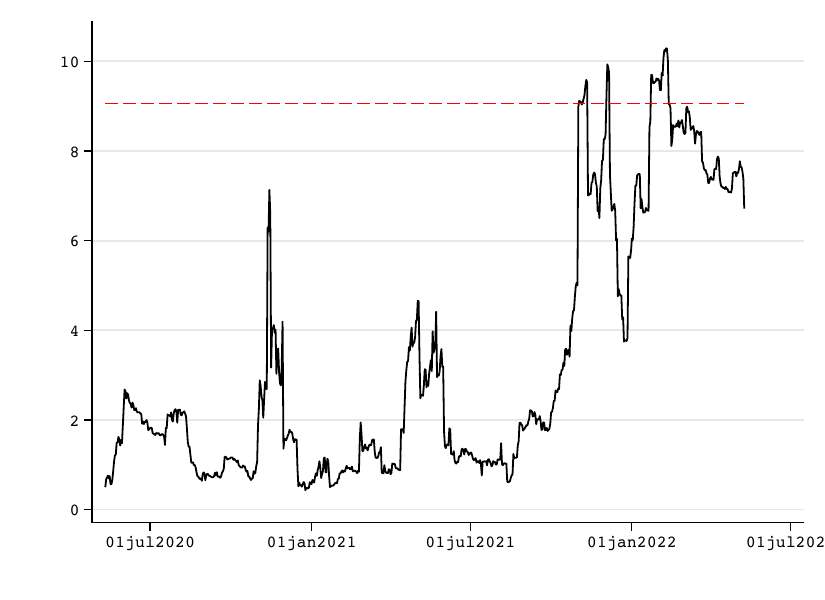}
    \caption{\fontsize{10}{12}\selectfont{Recursive Evolving-Heteroskedasticity.}}
  \end{subfigure}
  \caption{\fontsize{10}{12}\selectfont{Causality tests for attention to COVID-19 in France --> Bitcoin returns. \\ Notes: The test statistic sequence (---) is in black; the 5\% bootstrapped critical value sequence (--) is in red. The selected lag order is 1.}}
\end{figure}

\FloatBarrier

\subsection{Collectivistic Countries}

\hspace{1.4em}The results of the recursive evolving tests running from the attention to COVID-19 in India, China, Japan, and South Korea to Bitcoin returns are presented in Figures 5--8.\footnote{Test results using rolling window algorithms are reported in the Appendix.} Here, based on two Wald tests, all the countries show at least one causal episode.\footnote{The interpretations rely mainly on the results of the heteroskedastic-consistent test.} As seen in Figure 5, the public attention to COVID-19 in India exhibits two short episodes of causality around November 24, 2020 and January 12, 2022. Similarly, China has three episodes, the longest of which began on January 19, 2021 and lasted roughly 33 days. In contrast, people's attention to COVID-19 in Japan and South Korea is found to have much stronger impacts on Bitcoin returns, both in terms of duration and magnitude. In particular, the Japanese results are remarkable -- causality existed for almost the entire sampling period. While the longest causal event for South Korea is also more persistent than other countries, with the exception of Japan. Specifically, it took place around June 21, 2021 and lasted about 196 days. Overall, collectivistic countries here show a stronger causality with Bitcoin returns than individualistic countries.\par

\begin{figure}[thp]
  \begin{subfigure}[b]{0.5\textwidth}
  \centering
    \includegraphics[scale=0.5]{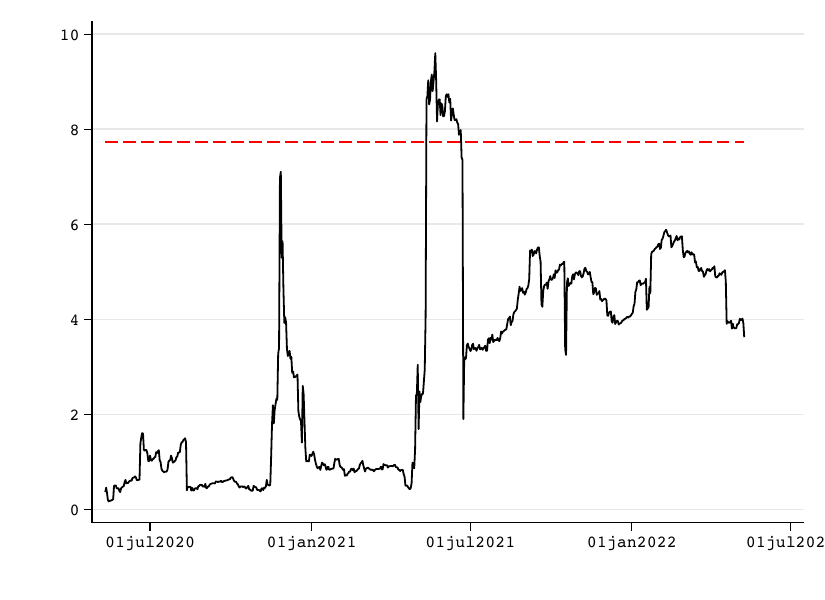}
    \caption{\fontsize{10}{12}\selectfont{Recursive Evolving-Homoskedasticity.}}
  \end{subfigure}
  \hfill
  \begin{subfigure}[b]{0.5\textwidth}
  \centering
    \includegraphics[scale=0.5]{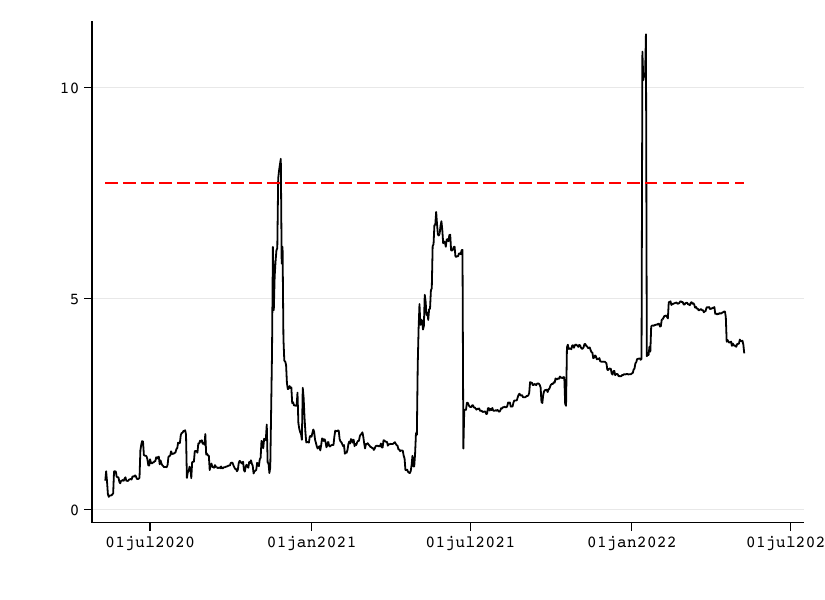}
    \caption{\fontsize{10}{12}\selectfont{Recursive Evolving-Heteroskedasticity.}}
  \end{subfigure}
  \caption{\fontsize{10}{12}\selectfont{Causality tests for attention to COVID-19 in India --> Bitcoin returns. \\ Notes: The test statistic sequence (---) is in black; the 5\% bootstrapped critical value sequence (--) is in red. The selected lag order is 1.}}
\end{figure}

\begin{figure}[thp]
  \begin{subfigure}[b]{0.5\textwidth}
  \centering
    \includegraphics[scale=0.5]{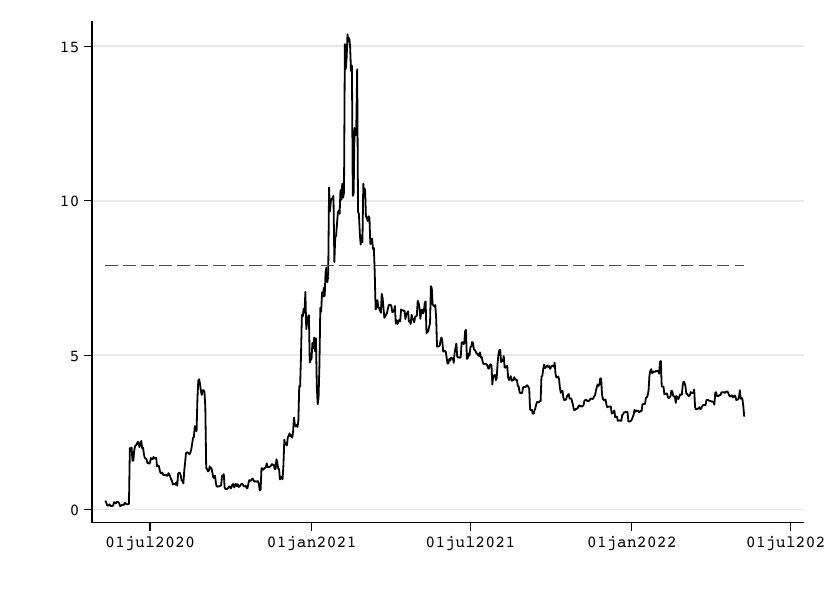}
    \caption{\fontsize{10}{12}\selectfont{Recursive Evolving-Homoskedasticity.}}
  \end{subfigure}
  \hfill
  \begin{subfigure}[b]{0.5\textwidth}
  \centering
    \includegraphics[scale=0.5]{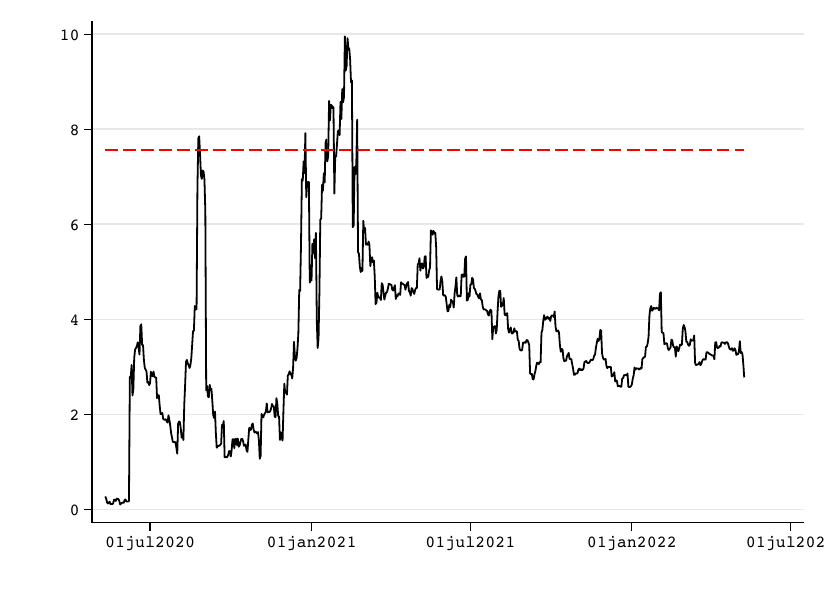}
    \caption{\fontsize{10}{12}\selectfont{Recursive Evolving-Heteroskedasticity.}}
  \end{subfigure}
  \caption{\fontsize{10}{12}\selectfont{Causality tests for attention to COVID-19 in China --> Bitcoin returns. \\ Notes: The test statistic sequence (---) is in black; the 5\% bootstrapped critical value sequence (--) is in red. The selected lag order is 1.}}
\end{figure}

\begin{figure}[thp]
  \begin{subfigure}[b]{0.5\textwidth}
  \centering
    \includegraphics[scale=0.5]{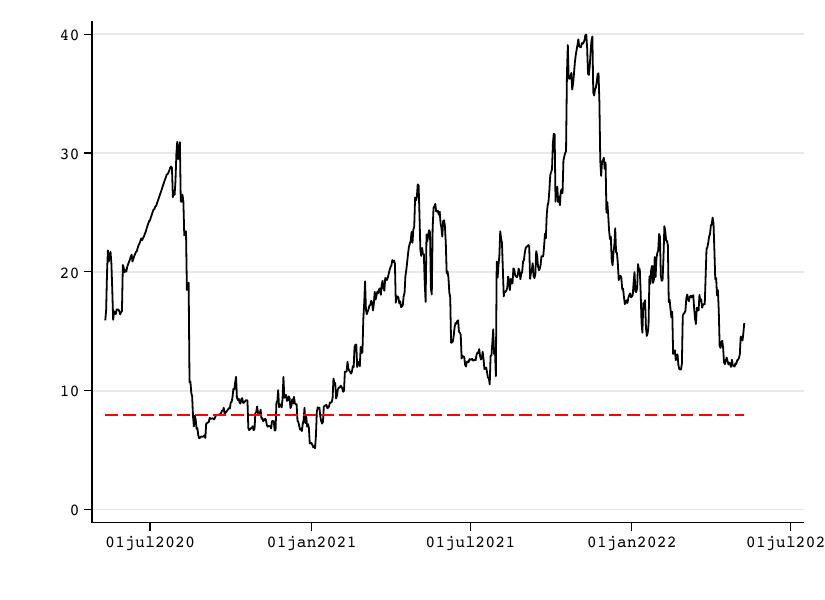}
    \caption{\fontsize{10}{12}\selectfont{Recursive Evolving-Homoskedasticity.}}
  \end{subfigure}
  \hfill
  \begin{subfigure}[b]{0.5\textwidth}
  \centering
    \includegraphics[scale=0.5]{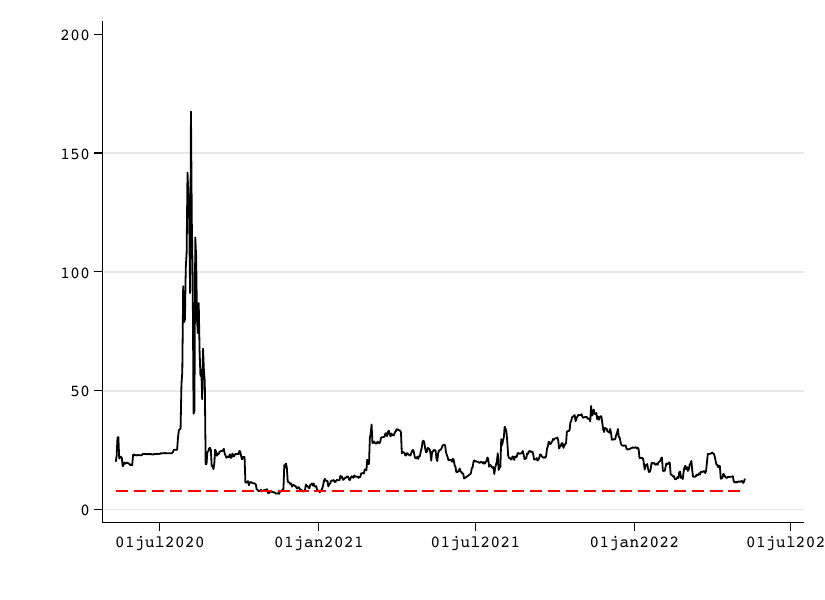}
    \caption{\fontsize{10}{12}\selectfont{Recursive Evolving-Heteroskedasticity.}}
  \end{subfigure}
  \caption{\fontsize{10}{12}\selectfont{Causality tests for attention to COVID-19 in the Japan --> Bitcoin returns. \\ Notes: The test statistic sequence (---) is in black; the 5\% bootstrapped critical value sequence (--) is in red. The selected lag order is 9.}}
\end{figure}

\begin{figure}[thp]
  \begin{subfigure}[b]{0.5\textwidth}
  \centering
    \includegraphics[scale=0.5]{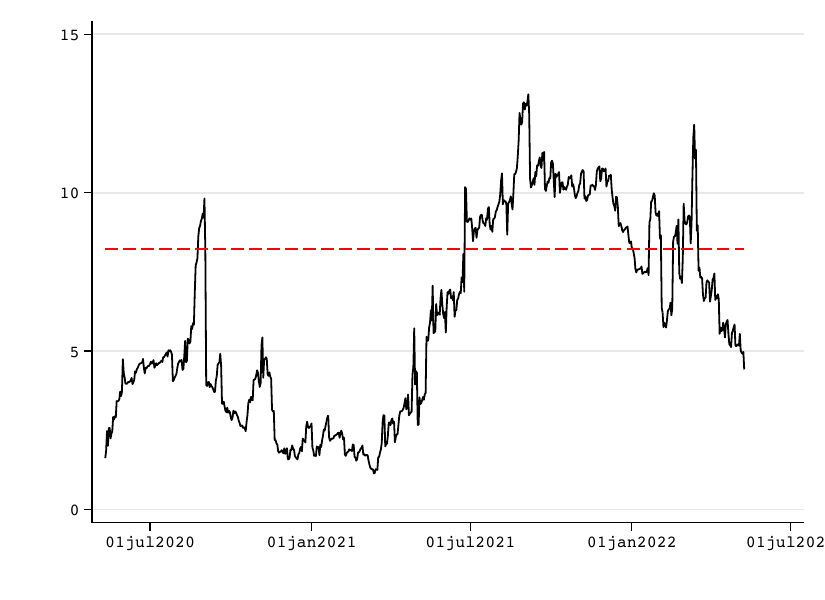}
    \caption{\fontsize{10}{12}\selectfont{Recursive Evolving-Homoskedasticity.}}
  \end{subfigure}
  \hfill
  \begin{subfigure}[b]{0.5\textwidth}
  \centering
    \includegraphics[scale=0.5]{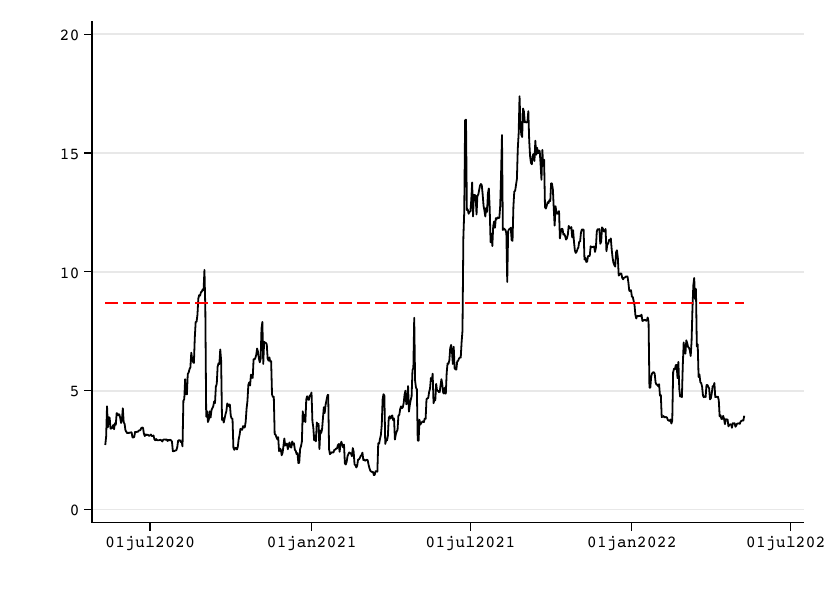}
    \caption{\fontsize{10}{12}\selectfont{Recursive Evolving-Heteroskedasticity.}}
  \end{subfigure}
  \caption{\fontsize{10}{12}\selectfont{Causality tests for attention to COVID-19 in South Korea --> Bitcoin returns. \\ Notes: The test statistic sequence (---) is in black; the 5\% bootstrapped critical value sequence (--) is in red. The selected lag order is 2.}}
\end{figure}

\FloatBarrier


\section{Conclusions}
\hspace{1.4em}This paper adopted two novel time-varying Granger causality tests to examine the relationships between regional attention to COVID-19 and Bitcoin returns. Eight large economies with a individualistic or collectivistic tradition were chosen for analyses. Collectivist countries have been found to have stronger causality ties with bitcoin returns than individualistic countries. The findings are consistent with earlier studies (e.g., \cites{Caporale2020, Chen2022}).  The paper adds ideas in a new light to the Bitcoin literature on individualism and COVID-19.\par



\section*{Acknowledgements}
This research did not receive any specific grant from funding agencies in the public, commercial, or not-for-profit sectors.

\section*{Appendix: Figures}

\begin{figure}[thp]
  \begin{subfigure}[b]{0.5\textwidth}
  \centering
    \includegraphics[scale=0.5]{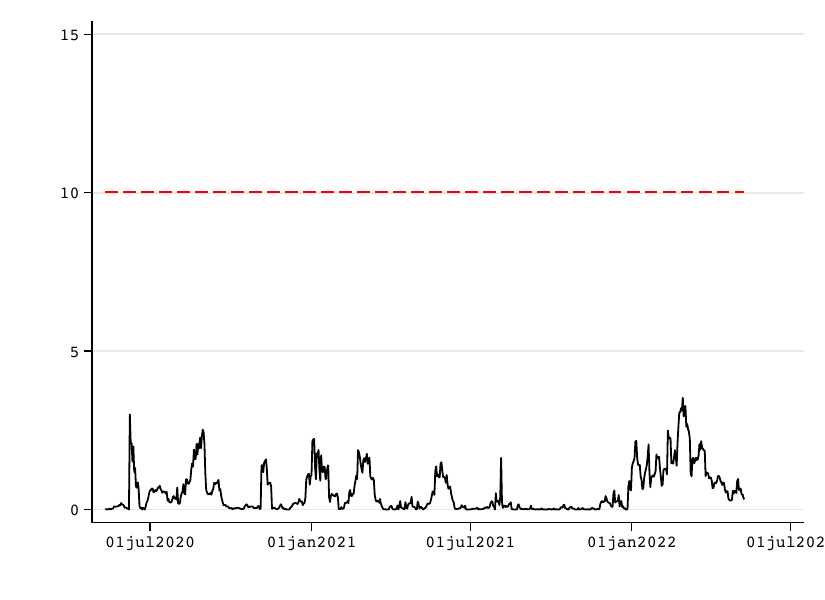}
    \caption{\fontsize{10}{12}\selectfont{Rolling-Homoskedasticity.}}
  \end{subfigure}
  \hfill
  \begin{subfigure}[b]{0.5\textwidth}
  \centering
    \includegraphics[scale=0.5]{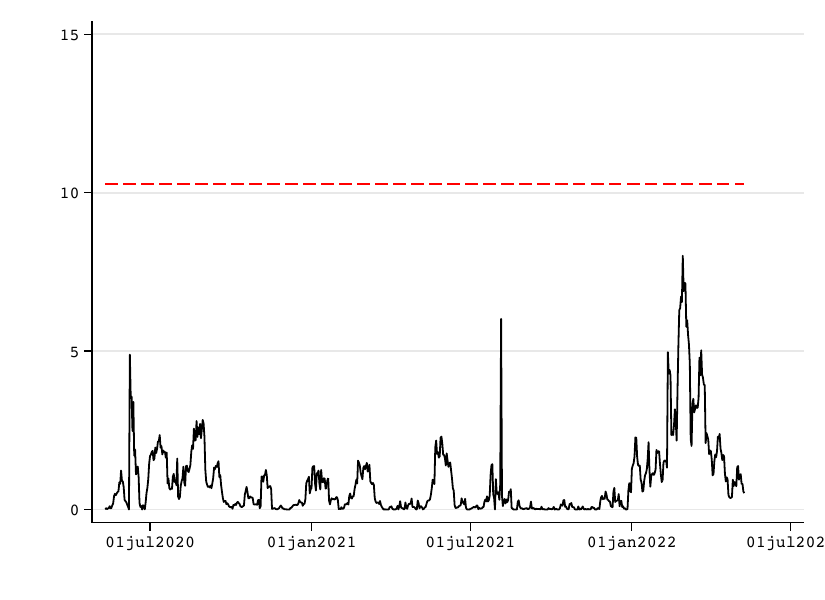}
    \caption{\fontsize{10}{12}\selectfont{Rolling-Heteroskedasticity.}}
  \end{subfigure}
  \caption{\fontsize{10}{12}\selectfont{Causality tests for attention to COVID-19 in the US --> Bitcoin returns. \\ Notes: The test statistic sequence (---) is in black; the 5\% bootstrapped critical value sequence (--) is in red. The selected lag order is 1.}}
\end{figure}

\begin{figure}[thp]
  \begin{subfigure}[b]{0.5\textwidth}
  \centering
    \includegraphics[scale=0.5]{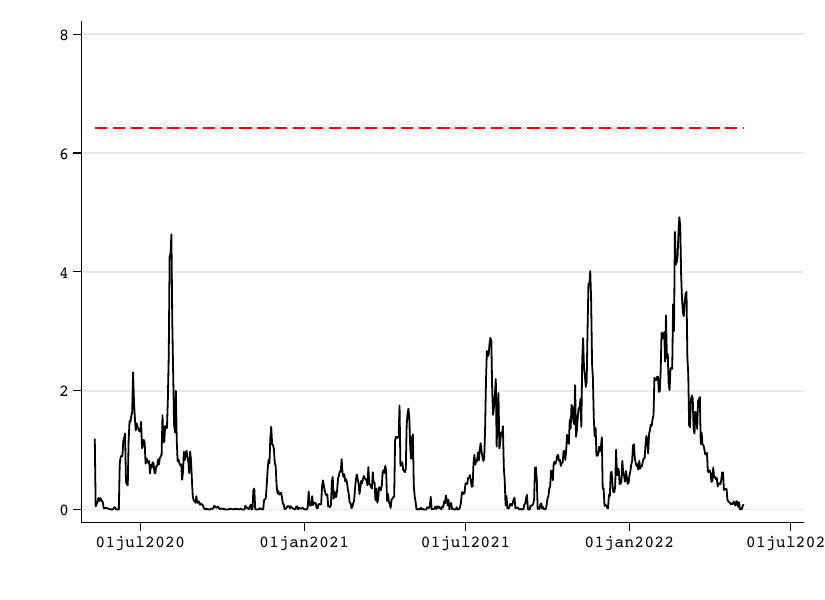}
    \caption{\fontsize{10}{12}\selectfont{Rolling-Homoskedasticity.}}
  \end{subfigure}
  \hfill
  \begin{subfigure}[b]{0.5\textwidth}
  \centering
    \includegraphics[scale=0.5]{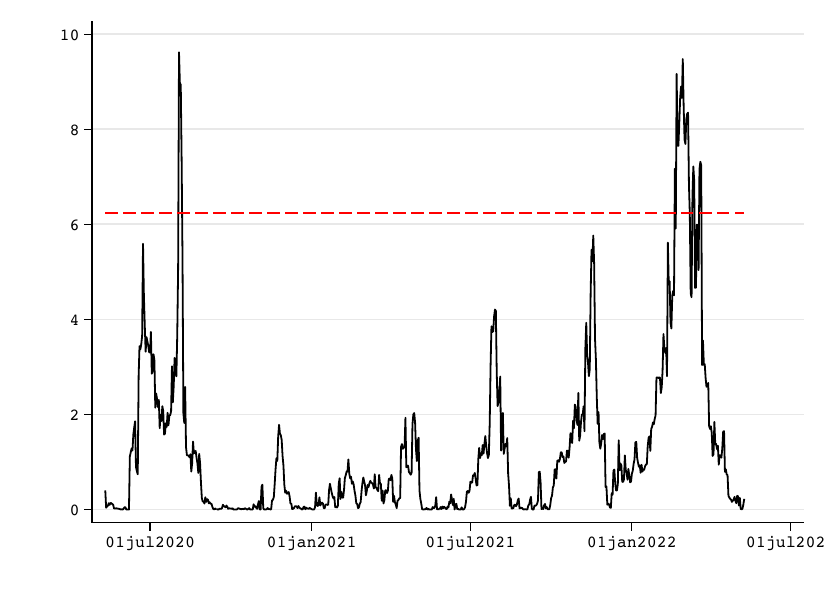}
    \caption{\fontsize{10}{12}\selectfont{Rolling-Heteroskedasticity.}}
  \end{subfigure}
  \caption{\fontsize{10}{12}\selectfont{Causality tests for attention to COVID-19 in Germany --> Bitcoin returns. \\ Notes: The test statistic sequence (---) is in black; the 5\% bootstrapped critical value sequence (--) is in red. The selected lag order is 1.}}
\end{figure}

\begin{figure}[thp]
  \begin{subfigure}[b]{0.5\textwidth}
  \centering
    \includegraphics[scale=0.5]{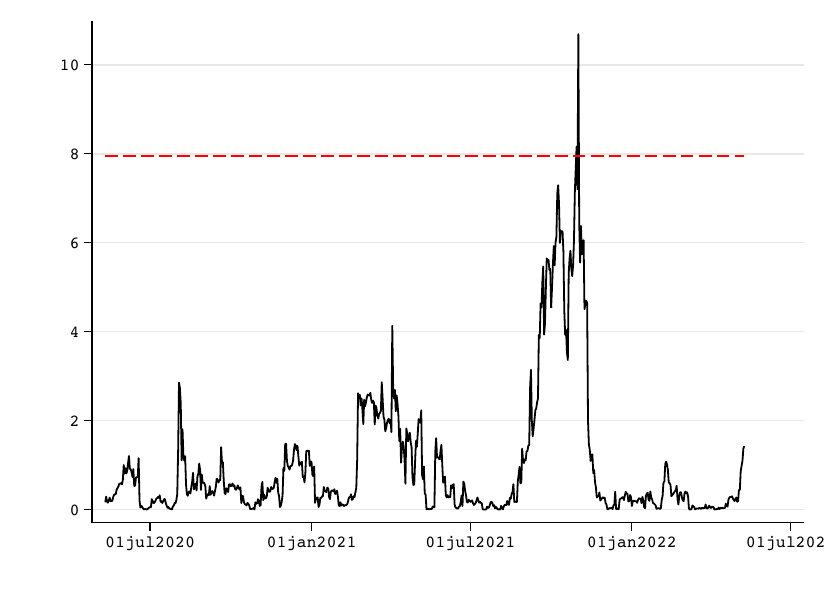}
    \caption{\fontsize{10}{12}\selectfont{Rolling-Homoskedasticity.}}
  \end{subfigure}
  \hfill
  \begin{subfigure}[b]{0.5\textwidth}
  \centering
    \includegraphics[scale=0.5]{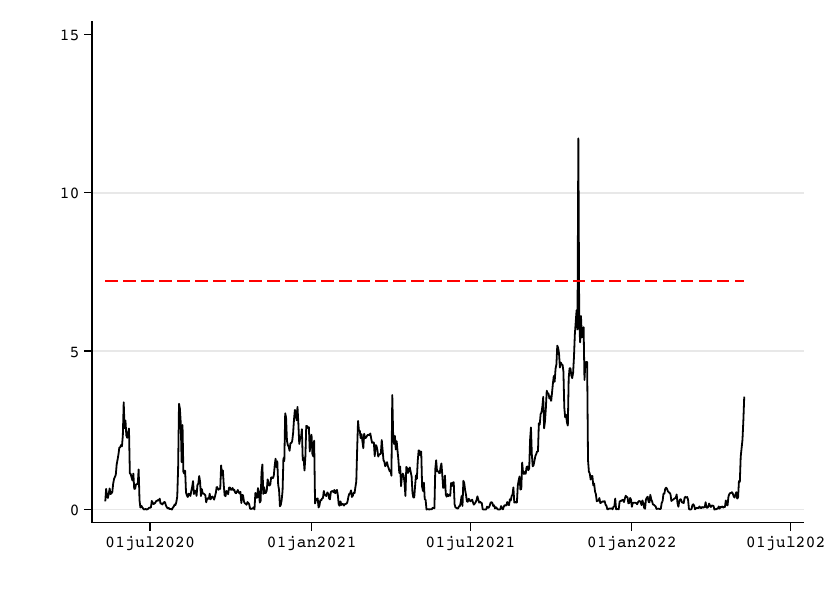}
    \caption{\fontsize{10}{12}\selectfont{Rolling-Heteroskedasticity.}}
  \end{subfigure}
  \caption{\fontsize{10}{12}\selectfont{Causality tests for attention to COVID-19 in the UK --> Bitcoin returns. \\ Notes: The test statistic sequence (---) is in black; the 5\% bootstrapped critical value sequence (--) is in red. The selected lag order is 1.}}
\end{figure}

\begin{figure}[thp]
  \begin{subfigure}[b]{0.5\textwidth}
  \centering
    \includegraphics[scale=0.5]{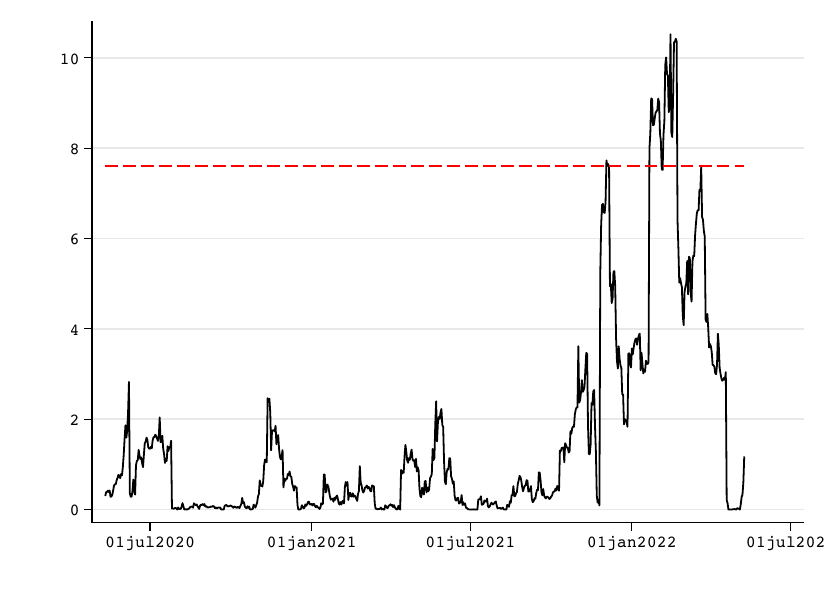}
    \caption{\fontsize{10}{12}\selectfont{Rolling-Homoskedasticity.}}
  \end{subfigure}
  \hfill
  \begin{subfigure}[b]{0.5\textwidth}
  \centering
    \includegraphics[scale=0.5]{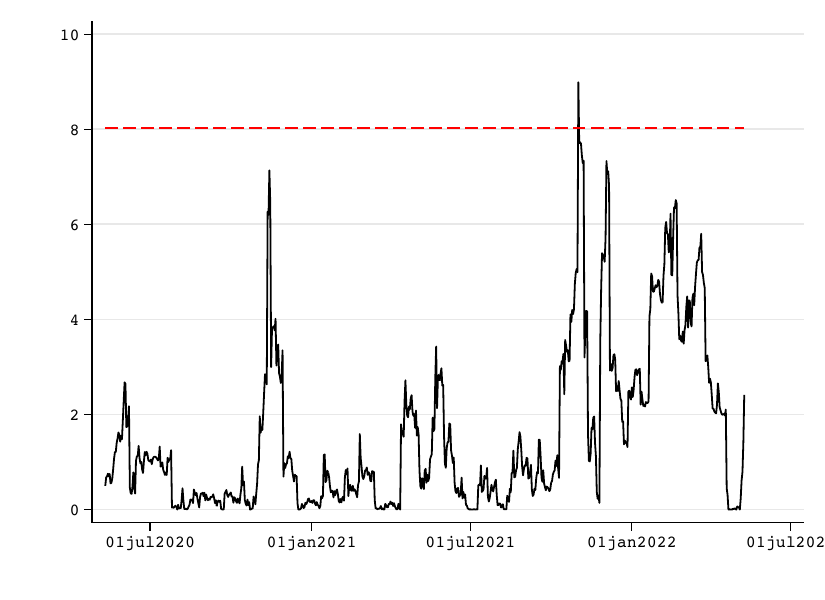}
    \caption{\fontsize{10}{12}\selectfont{Rolling-Heteroskedasticity.}}
  \end{subfigure}
  \caption{\fontsize{10}{12}\selectfont{Causality tests for attention to COVID-19 in France --> Bitcoin returns. \\ Notes: The test statistic sequence (---) is in black; the 5\% bootstrapped critical value sequence (--) is in red. The selected lag order is 1.}}
\end{figure}

\begin{figure}[thp]
  \begin{subfigure}[b]{0.5\textwidth}
  \centering
    \includegraphics[scale=0.5]{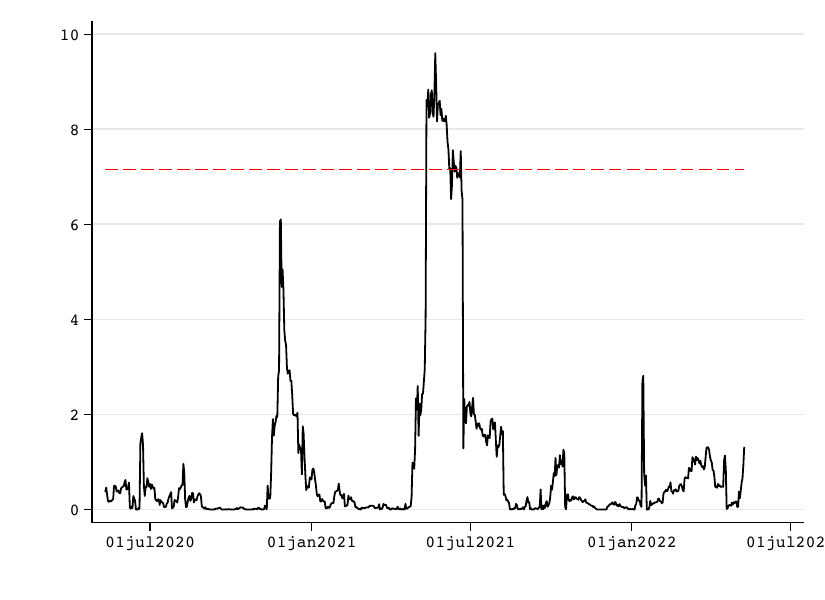}
    \caption{\fontsize{10}{12}\selectfont{Rolling-Homoskedasticity.}}
  \end{subfigure}
  \hfill
  \begin{subfigure}[b]{0.5\textwidth}
  \centering
    \includegraphics[scale=0.5]{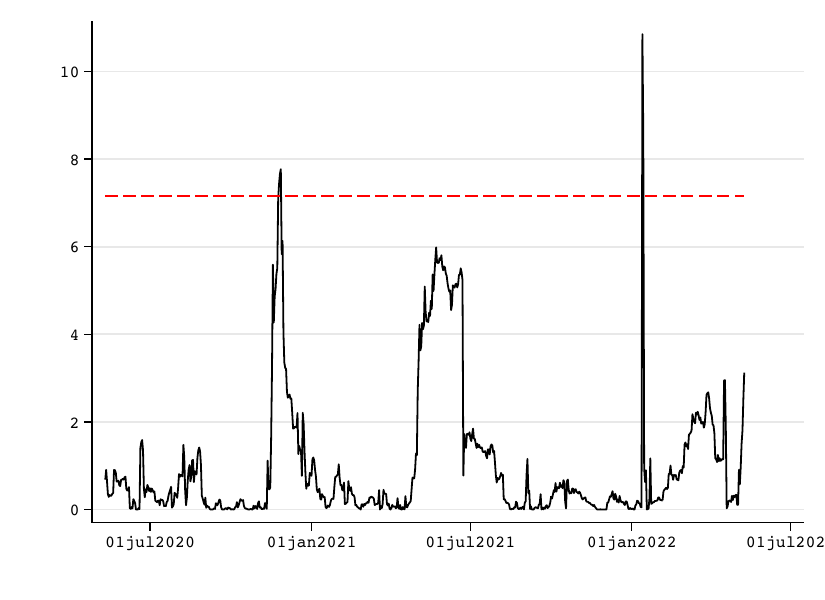}
    \caption{\fontsize{10}{12}\selectfont{Rolling--Heteroskedasticity.}}
  \end{subfigure}
  \caption{\fontsize{10}{12}\selectfont{Causality tests for attention to COVID-19 in India --> Bitcoin returns. \\ Notes: The test statistic sequence (---) is in black; the 5\% bootstrapped critical value sequence (--) is in red. The selected lag order is 1.}}
\end{figure}

\begin{figure}[thp]
  \begin{subfigure}[b]{0.5\textwidth}
  \centering
    \includegraphics[scale=0.5]{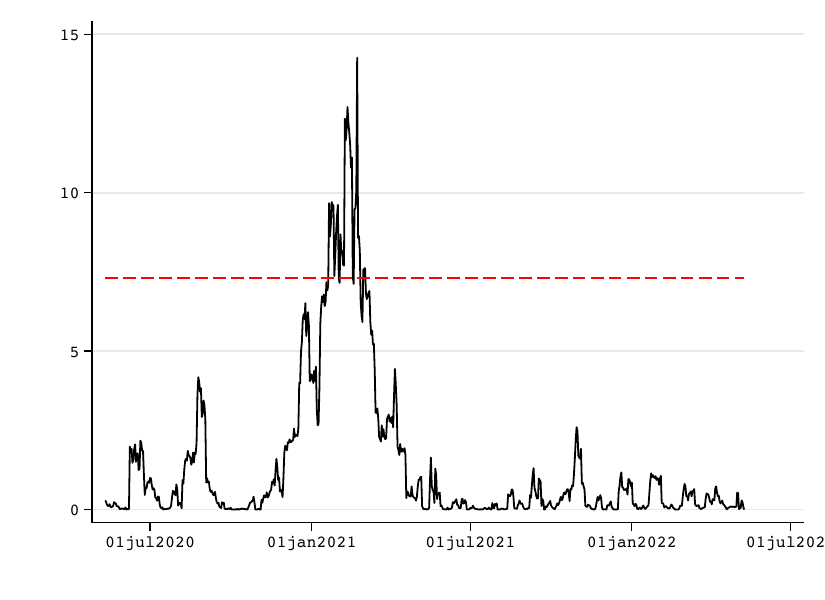}
    \caption{\fontsize{10}{12}\selectfont{Rolling-Homoskedasticity.}}
  \end{subfigure}
  \hfill
  \begin{subfigure}[b]{0.5\textwidth}
  \centering
    \includegraphics[scale=0.5]{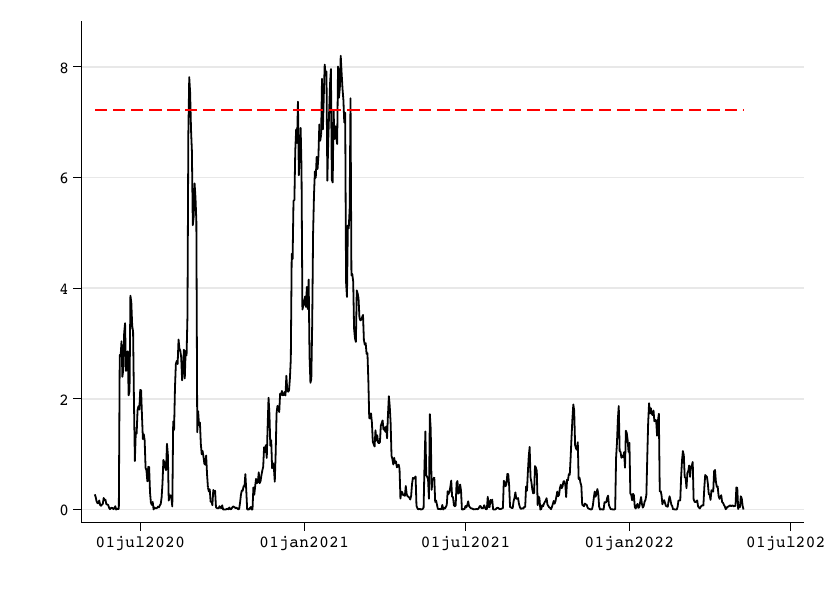}
    \caption{\fontsize{10}{12}\selectfont{Rolling-Heteroskedasticity.}}
  \end{subfigure}
  \caption{\fontsize{10}{12}\selectfont{Causality tests for attention to COVID-19 in China --> Bitcoin returns. \\ Notes: The test statistic sequence (---) is in black; the 5\% bootstrapped critical value sequence (--) is in red. The selected lag order is 1.}}
\end{figure}

\begin{figure}[thp]
  \begin{subfigure}[b]{0.5\textwidth}
  \centering
    \includegraphics[scale=0.5]{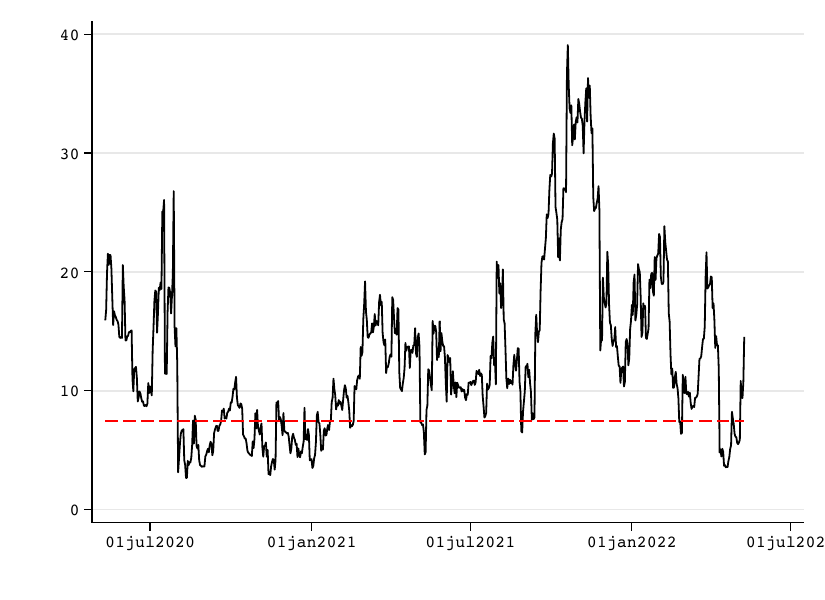}
    \caption{\fontsize{10}{12}\selectfont{Rolling-Homoskedasticity.}}
  \end{subfigure}
  \hfill
  \begin{subfigure}[b]{0.5\textwidth}
  \centering
    \includegraphics[scale=0.5]{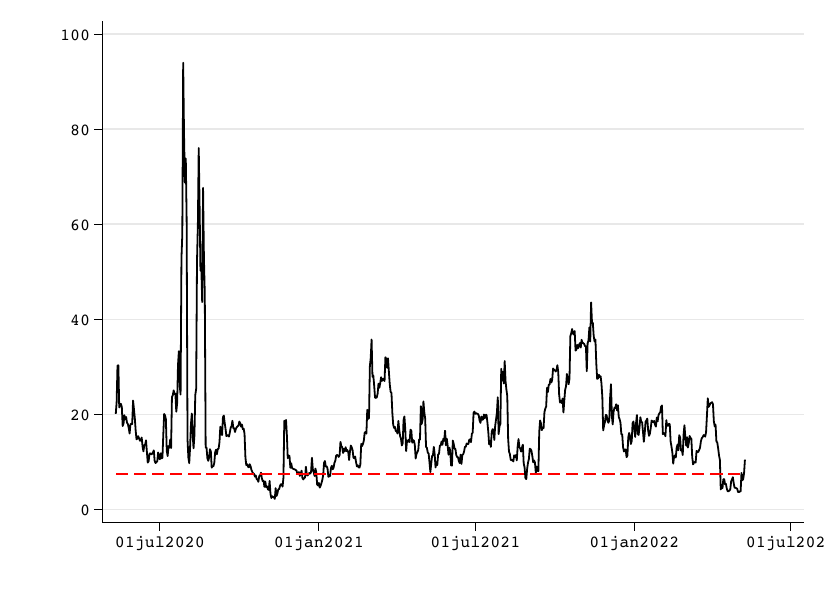}
    \caption{\fontsize{10}{12}\selectfont{Rolling-Heteroskedasticity.}}
  \end{subfigure}
  \caption{\fontsize{10}{12}\selectfont{Causality tests for attention to COVID-19 in the Japan --> Bitcoin returns. \\ Notes: The test statistic sequence (---) is in black; the 5\% bootstrapped critical value sequence (--) is in red. The selected lag order is 9.}}
\end{figure}

\begin{figure}[thp]
  \begin{subfigure}[b]{0.5\textwidth}
  \centering
    \includegraphics[scale=0.5]{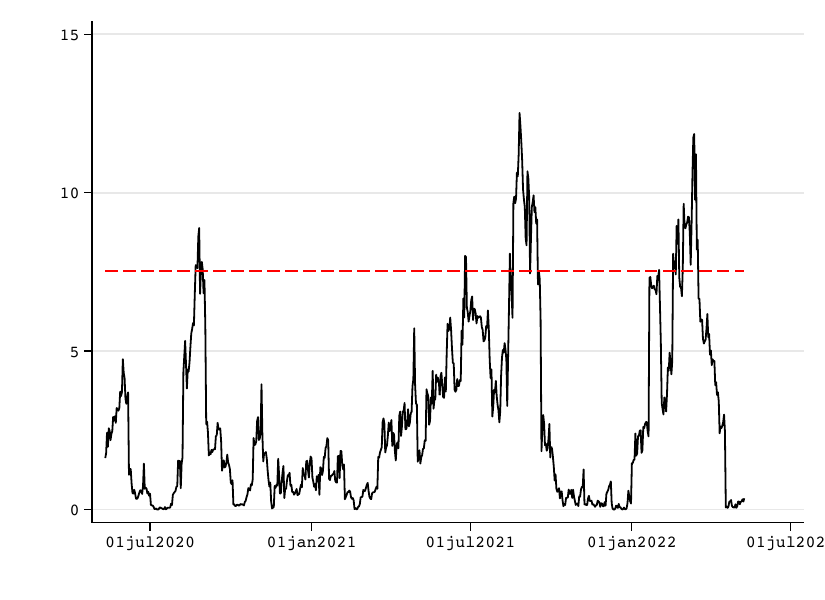}
    \caption{\fontsize{10}{12}\selectfont{Rolling-Homoskedasticity.}}
  \end{subfigure}
  \hfill
  \begin{subfigure}[b]{0.5\textwidth}
  \centering
    \includegraphics[scale=0.5]{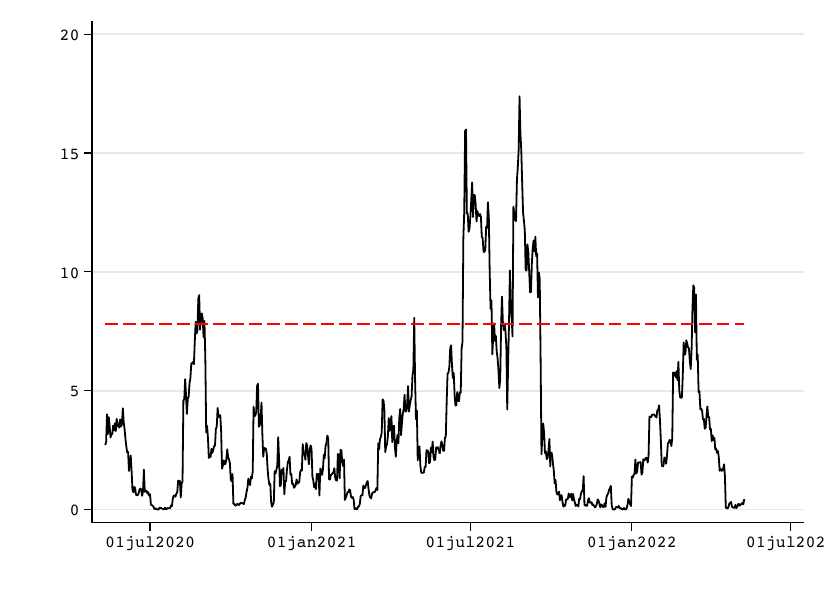}
    \caption{\fontsize{10}{12}\selectfont{Rolling-Heteroskedasticity.}}
  \end{subfigure}
  \caption{\fontsize{10}{12}\selectfont{Causality tests for attention to COVID-19 in South Korea --> Bitcoin returns. \\ Notes: The test statistic sequence (---) is in black; the 5\% bootstrapped critical value sequence (--) is in red. The selected lag order is 2.}}
\end{figure}

\clearpage 

\begingroup
\setstretch{0.9}
\setlength\bibitemsep{5pt}
\AtNextBibliography{\small}
\printbibliography
\endgroup

\end{document}